\documentclass[openacc]{rstransa}	%%%% where rstrans is the template name

\titlehead{Research}

\usepackage{color,multirow}            % to get colors
\definecolor{bla}{rgb}{0,0,0} % defines 'black'
\definecolor{red}{rgb}{1,0,0} % defines 'red'
\definecolor{blu}{rgb}{0,0,1} % defines 'blue'
\definecolor{gre}{rgb}{0,0.6,0} % defines 'green'
\definecolor{ora}{rgb}{1,0.5,0} % defines 'orange'
\definecolor{yel}{rgb}{0.7,0.7,0} % defines 'yellow'
\definecolor{pur}{rgb}{0.5,0,0.5} % defines 'purple'
\definecolor{cya}{rgb}{0,0.8,0.8} % defines 'cyan'

\newcommand\bla{\color{bla}}

\usepackage{comment} % for commenting large blocks of text
\usepackage{pdflscape}

\begin{document}

%%%% Article title to be placed here
%\title{Questions on the origins and ages of ring systems}
\title{Origins of rings in the solar system}

\author{%%%% Author details
B. Sicardy$^{1}$, 
M. El Moutamid$^{2}$, % Maryame, should you change your affiliation to SwRI?
S. Renner$^{1}$, 
R. Sfair$^{3,4,5}$ % Rafael has a double affiliation at LESIA
and
D. Souami$^{5,6}$ % LESIA & naXys (Namur)
}

%%%%%%%%% Insert author address here
\address{%
$^{1}$IMCCE, Observatoire de Paris, Universit\'e PSL, Sorbonne Universit\'e, Universit\'e de Lille, CNRS UMR 8028, 77 avenue Denfert-Rochereau 75014 Paris, France \\
$^{2}$Southwest Research Institute, Boulder CO 
%\bs{%%%
80302-5491
%}%%%
\\
$^{3}$ S\~ao Paulo State University (UNESP), School of Engineering and Sciences, 12516-410, Guaratinguet\'a, Brazil \\
$^{4}$ Institute of Astronomy and Astrophysics, University of T{\"u}bingen, Auf der Morgenstelle 10, 72076, T{\"u}bingen, Germany \\
$^{5}$LESIA, Observatoire de Paris, Universit\'e PSL, CNRS, Sorbonne Universit\'e, 5 place Jules Janssen, 92190 Meudon, France \\
$^{6}$ naXys, Department of Mathematics, University of Namur, Rue de Bruxelles 61, 5000 Namur, Belgium
}%

%%%% Subject entries to be placed here %%%%
\subject{Astrophysics}

%%%% Keyword entries to be placed here %%%%

%%%% Insert corresponding author and its email address}
\corres{Bruno Sicardy\\
\email{bruno.sicardy@obspm.fr}}

%%%% Abstract text to be placed here %%%%%%%%%%%%
\begin{abstract}
Until about a decade ago, ring systems were only known to exist around giant planets. Each one of the four giant planets harbours its own distinctive and unique system of rings and inner satellites. The past decade has been marked by the unexpected discoveries of dense rings around small objects of the outer solar system: the Centaur object Chariklo (and possibly Chiron), the dwarf planet Haumea and the trans-Neptunian object Quaoar. In the latter case, an additional surprise came from the fact that Quaoar's rings orbit well beyond the Roche limit of the body.
% (optical depth of order unity) % around the dwarf planet Quaoar \newline
%and around several small objects of the outer solar systems: the Centaur Chariklo (and possibly Chiron) and the trans-neptunian objects Haumea and Quaoar.
Here, we address the possible origins and evolution of these ring systems.
% existence, and stability  
%   Saturn's rings could stem from the disruption of a satellite that migrated inwards and got disrupted in the Roche zone of the planet.  \bla
%   The more elusive rings of Uranuss and Neptune may result from a stationary regime in which small satellites collide, form rings and accrete again into small objects.
%   Jupiter's dusty rings are the most tenuous of all, and are produced by the meteoroid bombardment of small satellites.
%   Rings around Centaurs could be the result of cometary material ejected from the surface of the bodies, forming a dense shell from which a ring emerged.
%Haumea's and Quaoar's rings may result from an impact that created a disk which later evolved into narrow rings.
%   Spin-orbit resonances between rings and small objects may be key to explain their confinements at specific radii.
%   Quaoar's rings orbit well beyond the Roche limit of the body, challenging the common  idea that rings should reside inside that limit, and bringing a fresh view on how rings form.
\end{abstract}
%%%%%%%%%%%%%%%%%%%%%%%%%%%

%%%%%%%%%% Insert the texts which can accomdate on firstpage in the tag "fmtext" %%%%%

\begin{fmtext}			% BS: I don't know what it is

\end{fmtext}

%%%%%%%%%%%%%%% End of first page %%%%%%%%%%%%%%%%%%%%%

\maketitle

%\noindent
%Color codes to use when implementing changes: \\
%\verb+\bs{corrections by Bruno}+ $\rightarrow$ \bs{corrections by Bruno} \\
%\verb+\mem{corrections by Maryame}+ $\rightarrow$ \mem{corrections by Maryame} \\
%\verb+\sr{corrections by Stefan}+ $\rightarrow$ \sr{corrections by Stefan} \\
%\verb+\rs{corrections by Rafael}+ $\rightarrow$ \rs{corrections by Rafael} \\
%\verb+\ds{corrections by Damya}+ $\rightarrow$ \ds{corrections by Damya} \\
%\verb+\pd{corrections still pending}+ $\rightarrow$ \pd{corrections still pending} \\

\vspace*{-5pt}
\section{Introduction}
\label{sec_intro}
%%%% Insert A head here

%[BS. 1/ I used the American English template, 2/ There are some problems w/ the layout of the bibliography part, e.g. ``4. Hubbard WBea,..."] \bla

Until the mid-1970s, planetary rings were only known to exist only around Saturn, these have been known since the 17$^{\rm th}$ century. It is only in the mid-1970s and early 1980s 
%\bs{%%%
that,
%}%%%
using ground-based stellar 
%\bs{%%%
occultations,
%%}%%%
rings were discovered in 1977 around Uranus \cite{elli77,mill77,bhat77} and in 1984 around Neptune \cite{hubb86}. In the meantime, Jupiter's tenuous rings were discovered on images taken during the {\it Voyager-1} flyby in 1979 \cite{smit79}.
%on March 4\textsuperscript{th}, 1979 \cite{smit79}.

Over the last decade, ground-based stellar occultations have revealed the unexpected presence of narrow and dense rings around small bodies of the outer solar system. Firstly, a ring around the Centaur Chariklo, which orbits between Saturn and Uranus, was discovered in 2013 \cite{brag14}. This was followed by the discovery in 2017 of a ring around the dwarf planet Haumea \cite{orti17}, which orbits at an average heliocentric distance of about 43~au. Most recently, a series of occultations observed between 2018 and 2022 revealed the presence of two rings around Quaoar, a large trans-Neptunian object that also orbits at an average heliocentric distance of 
%\bs{%%%
$\sim$43~AU
%%}%%%
\cite{morg23,pere23}. 

All the rings discussed here differ by several orders of magnitude in terms of radii and
%\bs{%%%
one order of magnitude in terms of heliocentric distances, while encircling
%%}%%%
very different objects. 
Moreover, dense and narrow structures embedded in a more diffuse halo have been detected around Chiron, yet another Centaur \cite{elli95,rupr15,sick20,orti23}.  
%\bs{%%%
The nature of this material is still debated. The sharp, time-varying features observed during occultations could indicate a ring system in formation, but no definitive answer to this hypothesis has yet been given.
%}%%%
%Two rings were first discovered in 2013 around Chariklo, a Centaur object orbiting between Saturn and Uranus \cite{brag14}, 
%then in 2017 a ring was seen around Haumea, a dwarf planet currently moving at about 50~AU from the Sun \cite{orti17}, while a series of occultations between 2018 en 2022 revealed the presence of two rings around Quaoar, a large trans-neptunian object (TNO) at heliocentric distance of about 43~AU \cite{morg23,pere23}.
%

%The nature of this material is still discussed. In any case, it evolves with time, and could be a ring system in formation. % So, in the rest of this paper, we will consider Chiron's material as a potential ring system, a position that may change in the future as more data are collected.

% It could be rings currently forming from a dust shell fed by cometary outbursts at Chiron's surface.  
% Thus, Chiron's system may be provide important clues on how rings form around small bodies, at least those with surface activity.

Nearly thirty small solar system bodies have been probed, thus far, using stellar occultations, with sufficient signal-to-noise-ratios to detect ring systems down to an optical depth limit of $\tau \sim$1\% and width of a few kilometers or more.
% {\color{red} est-ce qu'il ne faut pas donner une limite de l'epaisseur optique?}. 
While the statistical set of bodies is small, the ratio of three confirmed (possibly four) ring systems out of some 30 candidates 
%\bs{%%%
\cite{orti20}\footnote{See also \url{https://occultations.ct.utfpr.edu.br/results/}}
%}%%%
augurs the discovery of many more rings around small bodies in the near future
%\bs{%%%
\cite{sica24}.
%}%%%
%Having discovered four such systems in this modest sample augurs the discovery of many more rings around small bodies in the near future. 
%
%

This paper examines various scenarios that may explain the origin and evolution of these diverse  ring systems. Its scope is to better understand whether rings form by following the same fundamental physical processes,
or on the contrary, have very diverse origins.

\vspace*{-5pt}
\section{General considerations}
\label{sec_general}

For each of the aforementioned objects, Table~\ref{tab_parameters} summarizes the main parameters for the bodies and their ring systems (radii, masses, and Roche limits), while Figs.~\ref{fig_Jupiter_Saturn_Uranus_Neptune} and \ref{fig_Chariklo_Haumea_Quaoar} provide  schematic overviews of these systems.

%%%%%%%%%%%%%%%%%%%%%%%%%%%%%%%%%%%%%%%% TABLE tab_parameters (NB. THE ORIGINAL PORTRAIT VERSION IS AFTER "END{DOCUMENT} ") %%%%%%%%%%%%%%%%%%%%%%%%%%
%{\color{red}Damya: je propose de mettre e raon de Roche dans un sous-colonne planete, ca sera lpus lisible je pense. Il pourra etre immediatement compar\'e aux rayons.om pourra apres re=ajuster les largeurs des colonnes.}
%
\begin{landscape}
\begin{table}[!h]
 %\red
 \caption{Physical parameters of the bodies and their ring systems}
 \label{tab_parameters}
 \setlength{\tabcolsep}{1mm}
 \renewcommand{\arraystretch}{1.1}
% \begin{tabular}{lll|lll}
\begin{tabular}{p{2.47cm}p{1.71cm}p{1.30cm}p{1.57cm}p{2.71cm}|p{2.80cm}p{2.65cm}p{2.4cm}} 
 \hline
 \multicolumn{5}{c|}{Central object}  & \multicolumn{3}{c}{Principal rings}       \\
 \hline
 Name /ref. &  Mass & Radius &  Rotational & Ring mass ratio                        & Radius & Roche limit             & Maximum \\
            & (kg)  & (km)   &  period (h) & $\mu$ (Eq.~\ref{eq_mass_ratio}) & (km) & (km) (Eq.~\ref{eq_Roche_limit})  & optical depth $\tau$ \\
% & & & &  &  &  & \\
 \hline
 Jupiter\cite{depater18b}   &  $1.9 \times 10^{27}$   & 71,500 & 9.92& (0.05 - 5)$\times 10^{-21}$    & 122,000\,-\,129,000 & 210,000     & $8 \times 10^{-6}$ \\
%    &   &    &  &                     &             & \\
 Saturn\cite{cuzz18} & $5.7 \times 10^{26}$ & 60,300 & 10.7 & $3 \times 10^{-8}$  & 74,500 - 137,000    & 140,000     & $> 4$ \\
%  &     &    &      &                     &             & \\
 Uranus\cite{tyle86,nich18} &  $8.7 \times 10^{25}$ & 25,600 & 17.2& $2 \times 10^{-10}$  & 42,000 - 51,000  & 74,000      & $8$ \\
%   &     &    &       &                     &             & \\
 Neptune\cite{depater18a}  &  $1.0 \times 10^{26}$ & 24,800 & 16.1 & $< 10^{-10}$  & Adams/arcs: 62,900       & 78,000 &  0.003/0.1 \\
                           &                       &        &      &               & Le Verrier: 53,200  &        &  0.003 \\
 \hline
 Chariklo\cite{leiv17,morg21} &  $7.0 \times 10^{18}$  & 125  & 7.00 & $10^{-6}$ & C1R: 386 & 320  & 0.4     \\
                              &                        &      &      &           & C2R: 400 &      & $>0.1$ \\
 Chiron\cite{sick20,brag23,orti23}  &  $4.8  \times 10^{18}$     & 98 &  5.92  & $10^{-6}$    & 325, 425  & 290  & $0.4$   \\
%   &          &  &    &           &      & \\
 Haumea\cite{orti17} &  $4\times 10^{21}$ & 800 & 3.92 & $2 \times 10^{-7}$  & H1R: 2287 & 2660 & $0.1$   \\
%      &        &  &   &           &      &  \\
 Quaoar\cite{morg23,pere23} &  $1.2 \times 10^{21}$ & 545 & 17.7 & $< 2 \times 10^{-8}$  & Q1R: 4057 & 1780 & $0.25$ \\
                            &                       &     &      &                       & Q2R: 2520 &      & 0.01 \\
 \hline
 \end{tabular}
\vspace*{2pt}
\\
{\footnotesize {\bf Notes.}
For the giant planets, the radii are equatorial, for the small objects, they are volumetric equivalent. Only the principal rings of the giant planets are considered here. Except for Saturn and Uranus, the estimation of the mass ratio $\mu$ is very uncertain. The parameters of Chiron's system are very uncertain due to the lack of information on the nature of the detected material.}
\vspace*{-4pt}
\end{table}
\end{landscape}
%%%%%%%%%%%%%%%%%%%%%%%%%%%%%%%%%%%%%%%%%%%%%%%%%%%%%%%%%%%%%%%%%%

\subsection{Ring mass}

The rings considered here represent a very small fractional mass $\mu$ of the central body,
\begin{equation}
\mu= \frac{m_{\rm r}}{M} \ll 1\,,
\label{eq_mass_ratio}    
\end{equation}
where $m_{\rm r}$ is the mass of the rings and $M$ is the mass of the central body. 

For instance, Saturn's ring mass is estimated to 
%\bs{%%%
be
%}%%%
$1.5~\times~10^{19}$~kg, i.e. $\mu \sim  3 \times~10^{-8}$ \cite{iess19}. If accreted, and assuming a density of $\rho~=~400$~kg~m$^{-3}$ (typical of the small inner satellites of Saturn \cite{thom20}), this would create an icy satellite of radius $\sim$200~km only.
The mass of Uranus' rings is estimated to $2 \times 10^{16}$~kg, or $\mu \sim 2 \times 10^{-10}$ \cite{espo02,chia03}, corresponding to an icy satellite of radius $\sim$20~km.
As for Neptune's ring mass, no rough estimates, not even to an order of magnitude can be given. Nonetheless, data suggest that it is significantly smaller than that of Uranus' rings \cite{depater18a}.
An extreme case is that of Jupiter's rings, as this dusty tenuous feature might have a ring to planet mass-ratio value as small as $\mu \sim (0.05-5) \times 10^{-21}$ (see Section~\ref{sec_jupiter}).

In regards to the newly discovered rings around small bodies, for which most questions remain open, mass estimates are difficult because detailed observations or robust modellings of their dynamics are not yet available. Nonetheless, a rough estimate of 10$^{13}$~kg is given for Chariklo's rings \cite{pan16,sica18}, yielding $\mu \sim 10^{-6}$, corresponding to an icy satellite of radius $\sim$2~km. 
%\bs{%%%
For the other rings, we assume that their dense parts have all comparable surface densities, in order of magnitude. Thus, we apply a simple rule where the ring mass is proportional to the surface area of these dense parts.  
% that is mass Haumea's ring/mass Chariklo's ring = (70x2287)/(6x386)
Considering that Chariklo's and Haumea's rings have widths of 6~km and 70~km, respectively \cite{morg21,orti17}, and using the radii given in Table~\ref{tab_parameters},
%}%%% 
Haumea's ring mass can be estimated to roughly $\sim10^{15}$~kg. This corresponds to $\mu \sim 2 \times 10^{-7}$ and an icy satellite of about 5~km in radius.
As for Quaoar's rings, most of their mass should be contained in the arc structure (see Fig.~\ref{fig_Chariklo_Haumea_Quaoar}) that is embedded in the Q1R ring. This structure has a typical width of $\sim$5~km and extends at most over 70 deg of azimuth \cite{pere23}. Applying again a simple rule of proportionality, 
% that is mass Quaoar's ring/mass Chariklo's ring = (5x4057)/(6x386)x(70/360)
this yields a maximum ring mass of $2 \times 10^{13}$~kg, i.e. $\mu~<~2~\times~10^{-8}$, corresponding to an icy satellite of less than 2~km in radius, like for Chariklo.

\subsection{Ring evolution}

%{\color{red} transition violente - eventuellement mettre des sous-section?}\\

Whatever its origin is, a 
dense collisional system\footnote{
%\bs{%%%
A dense ring is defined as a system where the dynamics is dominated by collisions, i.e. where particles suffer a collision or so after a few revolutions. As the collision rate is $\sim 20\tau$ \cite{salo18}, this criterion corresponds to $\tau$ of a few percents.
%}%%%
}
surrounding a body rapidly flattens into a disk due to dissipation of energy and conservation of angular
%\bs{%%%
momentum
%}%%%
over short time scales comparable to the orbital period of the particles. However, its thickness remains finite, as the disk adjusts itself so that to be marginally stable against 
%\bs{%%%
local gravitational instabilities, also called Julian–Toomre gravity wakes \cite{juli66} (not to be confounded with the wakes excited by the passage a satellite near a ring edge).
%}%%% 
The disk reaches a state where the Toomre parameter \cite{toom64} 
\begin{equation}
Q= \frac{c \Omega}{\pi G \Sigma}
\label{eq_toomre}
\end{equation}
is close to unity, where $G$ is the gravitational constant and $c$ is the velocity dispersion in a Keplerian disk with surface density $\Sigma$ rotating at angular velocity $\Omega$.
The state $Q \sim 1$ expresses that the local instabilities maintain an equilibrium between  pressure and rotation (through $c$ and $\Omega$, respectively) that disperse the wakes, and gravity (through $\Sigma$) that tend to collapse them.
Simulations show for instance that the dense part of Saturn's rings have typically $Q \lesssim 2$
\cite{salo18}.
For a broad ring of radius $a$,
%\bs{%%%
$Q \sim 1$
%}%%%
implies that the ring thickness $h$ is related to the mass ratio $\mu$ by \cite{sica05,sica06}
\begin{equation}
\frac{h}{a} \sim \mu,
\label{eq_thickness}
\end{equation}
where ``broad" means here that $a$ is assumed to be a few times the radius of the body. 
%
%\newline
%by noting that the local velocity field around a ring particle looks very much like the local velocity field in Saturn's rings, so that its surface density is similar \cite{sica18}. Similar low values are obtained when considering the rings around Uranus, Neptune, Haumea and Quaoar. So, whatever their origins are, the rings need very few material compared to the body that they encircle.
%
% https://en.wikipedia.org/wiki/Rings_of_Jupiter -->

For Saturn, we have $\mu \sim 3 \times 10^{-8}$ and $a \sim 100,000$~km, so Eq.~\ref{eq_thickness} tells us that $h$ should be of the order of a few meters.
%that so that the equation above provides $h$ of some meters. 
This agrees with the local thickness of the rings derived from the behaviour of resonantly excited density waves \cite{cuzz18}. The thickness $h$ also represents a few times the size of the largest particles. This means that the icy material that composes Saturn's  main rings ended up in a myriad of particles where the biggest particles have meter-sized radii. 
%\bs{%%%
As all the ring particles considered here move in similar local relative velocity fields, the Julian–Toomre instabilities mentioned earlier should evolve to provide similar particle size distributions and thicknesses. Thus, Eq.~\ref{eq_thickness} should also apply, explaining why $\mu$ is expected to be small (Table~\ref{tab_parameters}).
%}%%%
% In other words, these rings involve a very amount of material compared to the body itself. 

At this point, we should be careful to distinguish between the origin of the material composing the rings, which may be ancient, and the origin of structures like gaps, sharp edges, narrow rings or ring-arcs, which may be recent.
In that respect, rings may be similar to geophysical features on Earth (mountains, oceans, polar caps, etc.) that
%\bs{%%%
continuously
%}%%% 
evolve over million-year time scales, while recycling material that has a much older origin.

As pointed out in the Introduction, rings are found around very diverse objects. Even inside the same class of objects (e.g. giant planets or Centaurs), the rings strongly differ from each other. 
Meanwhile, some similarities emerge. One of them is that out of the eight ring systems considered here, six (Uranus, Neptune, Chariklo, Haumea, Quaoar and Chiron) are composed of a few narrowly confined dense rings (or ``ringlets") with widths ranging from a few kilometers to a few tens of kilometers only. Thus, they occupy a small fraction of the area available in the equatorial plane of the body, while being separated by essentially "void" space if we neglect some tenuous dusty material. This suggests a ``convergent evolution", where initial broad disks of various origins end up at specific radial distances, through some generic confining mechanisms to be
%\bs{%%%
explained.
%}%%%

Saturn's and Jupiter's rings differ from this general picture, while being in some way the opposite of each other. Saturn's dense rings (Sect. \ref{sec_saturn}) are broadly distributed and practically fill the entire space between the planet and its Roche limit. Conversely, Jupiter's rings (Sect. \ref{sec_jupiter})  are extremely tenuous, without evidence of dense ringlets.

\subsection{Roche limit}

Another similarity shared by all the ring systems seen so far -- except one -- is that they reside 
inside or close to the Roche limit of their respective hosts. The Roche radius $a_{\rm R}$ can be estimated using the formula \cite{tisc13}
\begin{equation}
a_{\rm R} = \left( \frac{3M}{\gamma \rho} \right)^{1/3},
\label{eq_Roche_limit}
\end{equation}
where $\gamma$ is a numerical coefficient that depends on the shape and spin state of the body that is disrupted by tidal forces, and $\rho$ is its bulk density. We use here the value of $\gamma = 1.6$, relevant to bodies filling their lemon-shaped Roche lobes \cite{tisc13}.
%\bs{%%%
Concerning densities, we assume $\rho = 400$~kg~m$^{-3}$ for Saturn's satellites and the other bodies orbiting farther away. No information on the parent bodies (if any) of Jupiter's rings is available. However, underdense icy particles at that distance should sublimate quickly, so we assume here denser particles with $\rho = 860$~kg~m$^{-3}$ as determined for Amalthea \cite{and05}.
% From https://ssd.jpl.nasa.gov/sats/phys_par/
%}%%%

Table~\ref{tab_parameters} confirms the general trend that dense rings are indeed found inside or close to the Roche limit, as defined by Eq.~\ref{eq_Roche_limit} and using the values of $\gamma$ and $\rho$ given above.
%
%\bs{%%%
However, this quantities remain poorly known. For instance, depending on the cohesion of the material, values as low as $\gamma \sim 0.85$ could be considered \cite{tisc13}, which would increase $a_{\rm R}$ by about 25\%. Likewise, although a density $\rho= 400$~kg~m$^{-3}$ seems reasonable for the small Saturnian satellites \cite{thom20}, this quantity could vary for satellites around other bodies. In particular, Chariklo's rings could reside outside the Roche limit.
The clear exception is Quaoar, with rings that are located well beyond this limit for any reasonable values of $\gamma$ and $\rho$,
%}%%%
see Fig.~\ref{fig_Chariklo_Haumea_Quaoar}. 
Thus, while being a robust indicator of where rings should be, the Roche criterion is in no way an absolute rule. This brings a fresh view on how rings form and evolve, see Section~\ref{sec_quaoar}.

%%%%%%%%%%%%%%%%%%%%%%%%%%%%%%%%%%%%%%%%%%%%%%%%%%%%%%%%%%%%%%%%%%%%%%%%%%%%%%%%%%%%%%%%%%%%%%%%%
\begin{figure}[!t]
\centering
\includegraphics[width=9cm]{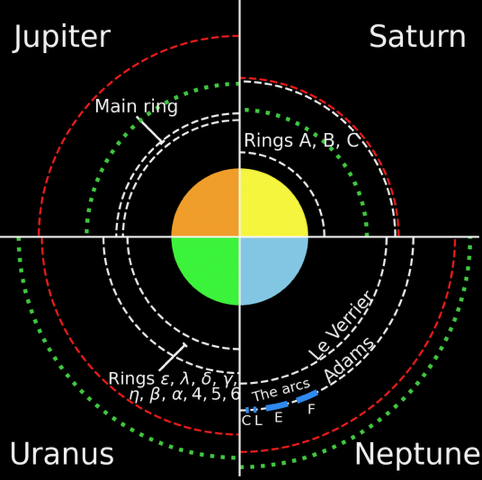}
\caption{%
A schematic view of the ring systems surrounding the giant planets, based on the values of Table~\ref{tab_parameters}, all scaled to the radius of the central planet. 
The white dashed lines delimit the principal rings of each planet, 
%\bs{%%%
the green dotted lines mark the corotation radii,
%}%%% 
while the red dashed lines indicate the Roche limits, using Eq.~\ref{eq_Roche_limit} and assuming ring particles with density 
%\bs{%%%
$\rho=400$~kg~m$^{-3}$ for Saturn, Uranus and Neptune, and $\rho=860$~kg~m$^{-3}$ for Jupiter, see text.
%}%%%
The letters C, L, E and F in the Neptune quadrant stand for the arcs Courage, Libert\'e, Egalit\'e and Fraternit\'e, respectively.
}%
\label{fig_Jupiter_Saturn_Uranus_Neptune}
\end{figure}

\subsection{Possible origins}

In view of the very diverse situations encountered in these ring systems, various scenarii can be evoked to explain the origins of rings. They can be included in different categories: 
\begin{itemize}
    \item
    the leftovers of a primordial disk that was present during the early formation stage of the central body, 
    \item 
    the result of a catastrophic impact that scooped material from the primary to form a disk from which satellites and rings emerged, 
    \item
    the outcome of the disruption of an object that would have entered the Roche limit of the body without damaging it. This object may be a passing body coming from infinity, or a satellite that migrated inwards, 
    \item
    the continuous erosion of small satellites by meteoroid bombardment that eject dust particles from their surfaces, creating very tenuous rings. Instead of bombardment, the ejected particles may also stem from geological activity (e.g., geysers) of the satellite.      
    \item 
    the result of collisions between small satellites that move near the body. These collisions may stem from the unstable nature of the satellite orbits or from the fact that the body had a close encounter with another object that destabilized the satellite(s). 
    \item 
    for active bodies, cometary jets may feed a dusty shell that flattens, thus forming a ring.
%    \item
%    \blu Rings around Iapetus and Rhea? \bla
\end{itemize}

%%%%%%%%%%%%%%%%%%%%%%%%%%%%%%%%%%%%%%%%%%%%%%%%%%%%%%%%%%%%%%%%%%%%%%%%%%%%%%%%%%%%%%%%%%%%%%%%%

%\bs{%%%
We will now briefly examine the specific ring systems that have been described above. A more detailed account of the origin of the rings around the giant planets and the Centaurs Chariklo and Chiron is given by \cite{char18}.
%}%%%

%\blu [We can be concise on the giant planets, as there are already several reviews on their possible origin, see e.g. \cite{char18}. More details on rings around small bodies can be given as this is new material] \bla

\vspace*{-5pt}
\section{Jupiter}
\label{sec_jupiter}

%%%%%%%%%%%%%%%%%%%%%%%%%%%%%%%%%%%%%%%%%%%%%%%%%%%%%%%%%%%%%%%%%%%%%%%%%%%%%%%%%%%%%%%%%%%%%%%%%%%%%%%%%%%%%%%%%%%%%%%%%
%\begin{comment}
% [Bruno] 
%The mass of Jupiter's rings (considering both dusty particle and larger parent bodies) is poorly constrained. 
%It should be in the range $10^{7}-10^{9}$~kg if only the dust is accounted for, and up to $10^{11}-10^{16}$~kg if %unseen parent bodies are considered \cite{burn04}. To give an idea of how little these masses are, we note that %compacted icy material of mass $10^{7}$~kg would fit into a sphere of radius fifteen meters only. 
%
%Jupiter: meteoroid bombardment of the small inner moons.  It has no dense component known. So the origin could be just a bit of meteoroid erosion on some of the small inner satellites Metis, Adrastea, Almathea and Thebe.
%\end{comment}
%%%%%%%%%%%%%%%%%%%%%%%%%%%%%%%%%%%%%%%%%%%%%%%%%%%%%%%%%%%%%%%%%%%%%%%%%%%%%%%%%%%%%%%%%%%%%%%%%%%%%%%%%%%%%%%%%%%%%%%%%

%[Explain clearly here that ``statements about the Roche limit, Toomre Q, and everything else that is relevant to all the other rings are not relevant to the tenuous ephemeral dusty rings of Jupiter" (Reviewer \#1).]%

%\bs{%%% by Rafael
As mentioned earlier, Jupiter's ring system stands apart from all other known ring systems in the solar 
system due to its unique, tenuous, and ephemeral nature. 
Unlike the dense, collisional rings found around other bodies, Jupiter's dusty rings 
do not conform to the typical considerations of Roche limits or Toomre stability criteria 
applicable to dense ring structures.
%}%%%
%Contrary to all other known rings in the solar system, Jupiter's rings consist of only
%tenuous dusty rings, with no dense component observed. 
The Jupiter ring system
comprises four main components, listed in increasing distance from the planet: the
Halo, Main, Amalthea Gossamer, and Thebe Gossamer rings. The innermost Halo ring
extends from about 92,000 to 122,500 kilometers from the planet's center and is a
thick, toroidal structure. Adjacent to this, the Main ring spans from 122,500 to
129,000 kilometers and is the brightest and narrowest part of the system. Beyond the
Main ring lies the Amalthea Gossamer ring, which stretches from 129,000 to 182,000
kilometers and is named after the moon Amalthea. The outermost Thebe Gossamer ring
extends from 182,000 to about 226,000 kilometers and is associated with the moon
Thebe \cite{depater15}. 

The optical depth of the rings is low, ranging from $\tau \sim 8 \times 10^{-6}$ to $ 10^{-9}$\cite{burn04, depater15}. 
The rings exhibit a strong forward-scattered signature, indicating a population dominated by
micrometric grains. This does not exclude the possibility of unseen macroscopic
bodies (a few meters in size), but the
%\bs{%%%
number
%}%%%
of these parent bodies is poorly
constrained. The evolution and lifetime of these dusty structures are determined by
the action of non-gravitational forces such as radiation pressure, electromagnetic
forces, and
%\bs{%%%
Poynting-Robertson
%}%%%
drag (P-R drag) \cite{burn01}. The expected
lifetime is $\lesssim 10^5$ years \cite{burn84, burn04}, much shorter than the
age of the solar system. Therefore, a continuous replenishment process must be in
place, with the impacts of interplanetary dust particles on the satellites and
collisions among larger bodies likely serving as sources for the rings. Examining each
ring individually, we can see how these processes contribute to their characteristics
and structures.

Amalthea and Thebe are located at the outer edge of the radial profile of their
respective rings, suggesting that meteoroid erosion of these satellites is likely the
source of these rings \cite{burn99}. Once ejected from the satellites, the particles
move inward due to
%\bs{%%%
P-R
%}%%%
drag, which explains the brightness peak of the rings
just inside the orbit of the satellites. Additionally, the vertical extent of both rings
matches the out-of-equatorial plane excursions of the inclined moons, corroborating
this model.

The edges of the Main ring are 
% {\color{red}related - DS- this verb is strange, should we say delimited or bound or ...} 
bound by the orbits of Metis and Adrastea \cite{burn04, throop04}, 
which supply particles to the ring through impacts in a
manner similar to the mechanics observed in the Gossamer rings. However, the
cross-sections of Metis and Adrastea are not large enough to be the only ring sources,
suggesting that an additional mechanism may be involved. There is an indication of
macroscopic material in the main ring \cite{throop04, porco03}, and impacts on these
bodies, as well as mutual collisions within this swarm of larger objects, that provide
additional dust.

The Halo ring is inside Metis' orbit and the inner radius of the ring is related to the $2:1$ Lorentz resonance.
%\bs{%%% By Rafael
A Lorentz resonance is defined when there is a commensurability between the orbital frequency and the periodicities of forcing frequencies of the Lorentz force acting on charged dust particles.
%}%%%
Dust generated by impacts in Metis, Adrastea, Amalthea, and
Thebe moves inwards and is vertically displaced by the $3:2$ Lorentz resonances,
causing the toroidal structure of the ring \cite{ockert99}.

\vspace*{-5pt}
\section{Saturn} \bla
\label{sec_saturn}

Most of the mass of Saturn's rings is contained in the main rings A, B, and C. Beyond Saturn's Roche limit, sparse dusty material exists,
%\bs{%%%
in particular the dusty F ring \cite{show92} orbiting between the satellites Prometheus and Pandora. 
This ring has proved to be highly dynamic, due to both the gravitational perturbations of the nearby  Prometheus and Pandora, and  ongoing collisions between putative small bodies in its core \cite{murr18}
Other dusty rings are seen
%}%%%
near the orbits of small satellites like Janus, Anthe, Methone, and Phoebe, with very low optical depths \cite{hedm18}. Similar to Jupiter's rings, Saturn's rings are sustained by impacts from micrometeoroids originating from these satellites. In contrast, the faint E ring is replenished by icy particles ejected from geysers on Enceladus' surface.

Several theories are proposed to explain the formation of Saturn's main rings A, B and C,
%\bs{%%%
see the synthesis presented by \cite{ida19}.
%}%%%
One hypothesis suggests they formed through the tidal disruption of a passing Centaur object or a large comet within Saturn's Roche limit
%\bs{%%%
\cite{done91,hyod17a,hyod17b},
%}%%%
akin to models proposed for the Moon's formation \cite{Opik1972},
%\bs{%%%
and as was observed during the tidal breakup of Comet Shoemaker–Levy 9 by Jupiter in 1992 \cite{asph96,movs12}.
%}%%%
%
This scenario posits that a comet passing near Saturn could have been torn apart by tidal forces, with fragments becoming weakly bound to Saturn. 
%\bs{%%%
A variant of this scenario would be the disruption of a saturnian satellite by an incoming comet that stripped the satellite mantle which was then pushed inward and formed an icy ring \cite{dubi19}.
However, the general problem with these scenarios is their low probability of occurrence.
%}%%%

%\bs{%%%
This motivated alternative models that
%}%%%
involved the tidal breakup of a large Saturnian satellite that migrated inward before being disrupted within the Roche zone \cite{canu10}. This process, referred to as peeling, describes how a satellite's less dense layers are stripped away by tidal forces as it moves inward, leaving behind debris.
In the same vein, the hypothetical "Chrysalis" satellite scenario, recently explored by \cite{wisd22}, suggests a destabilized moon as a potential explanation for the rings' formation. 
%
%\bs{%%%
More precisely, backward numerical integrations by \cite{wisd22} show that the current Saturn's obliquity is consistent with the destruction of this hypothetical moon when it approached Saturn to form the rings some 100 million years ago, supporting the idea of a young system.
% This model contributes to the debate over the rings' age, with some estimates suggests that Saturn's rings are young % possibly a few hundred million years old, while others propose an age of 4.5 billion years. 
This scenario was actually developed to explain the young-age estimates based on observations of non-icy materials in the rings, whose presence is caused by a relatively recent contamination due to ongoing micrometeoroid impacts, as discussed now.
%}%%% 
% The age debate is further complicated by uncertainties in understanding the rings' evolutionary history. 
%
%\pd{Bruno to Maryame? All?: I don't know how to respond to Reviewer \#1's comment: ``I don’t understand how destruction of a moon close to Saturn to form the rings being a rare event supports the young age for the rings". Idem with Reviewer \#2's comment: ``This sentence isn't clear. I think the authors are referring to a scenario like the one proposed by Dubinski (Icarus 321, 291, 2019), in which the impact of a Centaur into a Mimas-like satellite makes the rings. Like tidal disruption of a Centaur, this is unlikely to have occurred recently, so that *motivates* other ideas such as Chrysalis."}
%
%[See changes in red above].

%\pd{Bruno to Maryame? All?: I do not know how to respond to Reviewer \#2's comment: ``It would be better to place references \cite{Zhang2017a} and \cite{Zhang2017b} in the previous paragraph". Moreover, the Reviewer \#2 also asks to quote \cite{lint24} here.}
%
%\bs{%%%
The ``contamination method" consists in measuring the (small) abundance of non-icy material in Saturn's rings, and confront it to the contamination rate by external dust, mainly coming the Kuiper Belt region \cite{kemp23}.
%}%%%
%
In this context, \cite{Zhang2017a,Zhang2017b} present results obtained from the Cassini spacecraft microwave radiometry and VLA interferometer. They report that Saturn's C ring particles are 70$\%$-75$\%$ porous and contain 1-2$\%$ silicates, with higher concentrations (6$\%$-11$\%$) in the middle
%\bs{%%%
of C ring.
%}%%%
%if mixed with water ice. 
They propose models where silicates may exist as chunks rather than fine powder, influencing opacity. Their preferred scenario suggests continuous pollution of the C ring by meteoroid impacts over 15-90 million years, with additional contamination 10-20 million years ago from a disrupted Centaur. In the B ring, silicate fractions range from 0.3$\%$ - 0.5$\%$, correlating with optical depth and up to 90$\%$ porosity. The A ring interior to the Encke gap shows silicate fractions of $\sim$0.2$\%$ - 0.3$\%$ with porosities from 55$\%$ - 90$\%$. The Cassini Division, similar to most C ring regions, exhibits 1$\%$ - 2$\%$ silicates with high porosity (>90$\%$). Overall, pollution exposure times for the A and B rings and the Cassini Division suggest an age of $ \lesssim$ 150 million years, aligning with theories that Saturn's rings originated from a recent icy moon breakup. 
%
%\bs{%%%
Complementary data from the Cassini Cosmic Dust Analyzer instrument (CDA) support this finding, as a more accurate incoming flux of dust has been assessed, resulting in a young age (100 to 400 million years) for Saturn's rings \cite{kemp23}. Meanwhile, the CDA results also suggest that the infalling grains may have their composition altered through impact-triggered chemistry \cite{lint24}, which can make results difficult to interpret.
%}%%%

%\pd{Bruno to Maryame? All?: I don't know how to respond to Reviewer \#1's comment: ``This section is missing a discussion of the implications of ballistic transport modeling on the age of Saturn’s rings".}
%\bs{%%%
Independently of observations, dynamical models proposed by \cite{cuzz90,estr23} that include mass loading and ballistic transport caused by micrometeoroid bombardment yields the same general conclusion, that is a young ring system with an age of a few hundred million years only.
%}%%%

%BS. Stefan suggests to mention \cite{crid19} somewhere here, I agree w/ him]
%--> Maryame: Done. Great suggestion! Thanks Stefan. ]
Proponents of an older age for the rings argue that they were once more massive and rockier, evolving over billions of years to their current icy composition through processes that removed rocky materials while preserving the ice. \cite{crid19} highlight unresolved issues and propose an interpretation consistent with the rings being as old as the Solar System. 
They referenced Cassini measurements showing ongoing matter loss from the rings to Saturn, a process largely unexplained but possibly responsible for gradually brightening the rings by cleaning their ice content. They argued that the rings' mass aligns with expectations of how ancient rings evolve: spreading over time, creating moons at their outer edge, and losing mass inward to Saturn. Even if originally massive, primordial rings would naturally evolve to their current mass
%\bs{%%%
\cite{salm10}. These studies suggest
%}%%% 
that if the rings were young, their current mass would be coincidental, which challenges the young-rings hypothesis.
%\bs{%%%
Later, \cite{crid21} also argued that some cleaning mechanisms could be at work in Saturn's rings, so that the rings may appear to be young while being actually old. This scenario is investigated further by \cite{espo24}, who argue that the non-icy pollutant may be destroyed over time, so that its abundance only provides a \it lower \rm limit of 400 to 1600 million years for Saturn's rings.
%}%%%

\vspace*{-5pt}
\section{Uranus} \bla
\label{sec_uranus}

%\begin{comment}
%[Bruno]
%Continuous collisions between small satellites (``moonlets") that move on unstable orbits \cite{dunc97}, leading to permanent collisions and re-accretion, i.e. continuous renewal of the ``ring landscape". 
%Cite \cite{hedm21}
%\end{comment}

The Uranian ring system, the second ring system to be discovered, is composed
of nine narrow and dense main rings \cite{elli77, mill77, bhat77}.
These rings, named 6, 5, 4, $\alpha$, $\beta$, $\eta$, $\gamma$, $\delta$, and
%\bs{%%%
$\epsilon$,
%}%%%
sorted by increasing distance to the planet, 
have widths of $<10$~km, with the exception of the outermost
$\epsilon$ ring, which has a width ranging from $20$ to $100$~km \cite{french86,
nich18}. Additionally, high phase angle images from Voyager II revealed
dust bands dispersed among the main rings \cite{smith86}, as well as a series
of narrow dusty features \cite{hedm21}.
Completing the Uranian ring family, images by the Hubble Space Telescope revealed 
the existence of a secondary ring system orbiting outside the main rings, 
comprising two dust components: the $\nu$ and $\mu$ rings \cite{showalter06}.
%\bs{%%% by Rafael
The innermost moons, Cordelia and Ophelia, shepherd the $\epsilon$ ring; 
the tenuous $\nu$ and $\mu$ rings and eleven other small satellites are found within the 
orbit of Miranda.
%}%%%
%The rings of Uranus maintain a close relationship with the small satellites of
%the Portia family, Puck, Mab, Perdita, and the two shepherd moons of the
%$\epsilon$ ring, Cordelia and Ophelia. These satellites are only tens of
%kilometers in size \cite{showalter06} and orbit interior to Miranda.

In the secondary ring system, the majority of particles in both rings are lost
due to solar radiation pressure in less than 1,000 years \cite{sfair09}. The
$\mu$ ring is likely generated by meteoroid collisions with Mab's surface,
explaining its distinct triangular profile aligned with the satellite's orbit
and its blue color \cite{sfair12}. The $\nu$ ring also has a triangular profile
and is bounded by Portia and Rosalind; however, contrary to the $\mu$ ring,
there is no satellite associated with its profile peak. It is postulated that
mutual collisions between macroscopic bodies embedded in the region are the
source for this ring, which could explain its red color \cite{depater06}.

Estimates of disruption due to impacts of cometary objects suggest that the
small moons of Uranus are unlikely to be primordial \cite{colwell92, colwell93,
zahnle03}. Instead, on a timescale of 
%\bs{%%% by Rafael
$< 10^9$
%}%%% 
years, a catastrophic impact 
occurs, leading to a collisional cascade
%\bs{%%% by Rafael
\cite{nesv23}.
%}%%%
These fragments exist in a dynamic environment with complex interactions \cite{dunc97, quillen14, cuk22}, and
planetary tidal forces prevent efficient re-accretion. Thus, the main rings may
result from a continuous process of collisional fragmentation and their subsequent
confinements.

\vspace*{-5pt}
\section{Neptune}
\label{sec_neptune}

The origin of Neptune's rings may follow the same general lines as for Uranus', except that no dense continuous rings were formed, while distinctive arc features are maintained over decadal time scales.

The Neptune ring system essentially consists of five tenuous rings, two broad (Galle and Lassell) and three narrow (Adams, Le Verrier and Arago), interspersed among the four innermost satellites Galatea, Despina, Thalassa, and Naiad.
Neptune's rings have less sharp edges and are less dense than the narrow rings of Uranus. They contain more dust, reminiscent of what is observed in Saturn's F ring in the case of the Adams and Le Verrier rings \cite{porco95}.
They are probably young, and probably result from the collisional fragmentation of small inner satellites. 
% on unstable orbits.
%
%\bs{%%%
More precisely,
%}%%%
because the synchronous orbit lies beyond the ring system
%\bs{%%%
(Fig.~\ref{fig_Jupiter_Saturn_Uranus_Neptune}),
%}%%%
the satellites migrate inwards under the effect of tidal forces induced by the planet. Therefore a cycle of
%\bs{%%%
destructive collisions and
%}%%%
re-accretion may be at work, modifying the structure of the ring system over time \cite{char18}. 

The most intriguing feature around Neptune is the ring arcs ($\tau \sim 0.1$), embedded in the more diffuse Adams ring ($\tau \sim 0.003$).
Four arcs (Courage, Libert\'e, Egalit\'e and Fraternit\'e, see Fig.~\ref{fig_Jupiter_Saturn_Uranus_Neptune}) were identified during the NASA/Voyager flyby in 1989 \cite{smith89,porco91}. As of today, the two leading arcs Courage, Libert\'e have faded away, whereas the other two appear to be stable \cite{depater18a}.  

Confinement mechanisms (involving mean motion resonances) are required, since the differential Keplerian motion should destroy the arcs in a few months. 
One possibility is that the satellite Galatea, located at 980 km inside the ring, ensures the azimuthal and radial confinement of the arcs \cite{porco91}, via the simultaneous actions of corotation and Lindblad resonances \cite{gold86}. Another solution is to consider a few co-orbital moonlets in relative equilibrium, undetected so far, generalizing the classical Lagrange $L_4$/$L_5$ configurations \cite{renner14}. Finally, another alternative is to consider a three-body resonance between the arcs and the nearby satellites Galatea and Larissa~\cite{showalter17}.

The breakup of a parent satellite or the accretion of ring material within the Roche zone of the planet (located outside the Adams ring, see Fig.~\ref{fig_Jupiter_Saturn_Uranus_Neptune})
could lead to the formation of this unique arc system.  
Perhaps the rapid evolution of the arcs over the decadal time scale results from a recent cometary impact onto a larger ring body \cite{showalter20}. 
%
%If Galatea and/or the co-orbital configurations play a role the timescales are longer \cite{renner14,depater18a}.

\vspace*{-5pt}
\section{Rings around small bodies: resonances}
\label{sec_resonances}

\begin{figure}[!t]
\centering
\includegraphics[width=12cm]{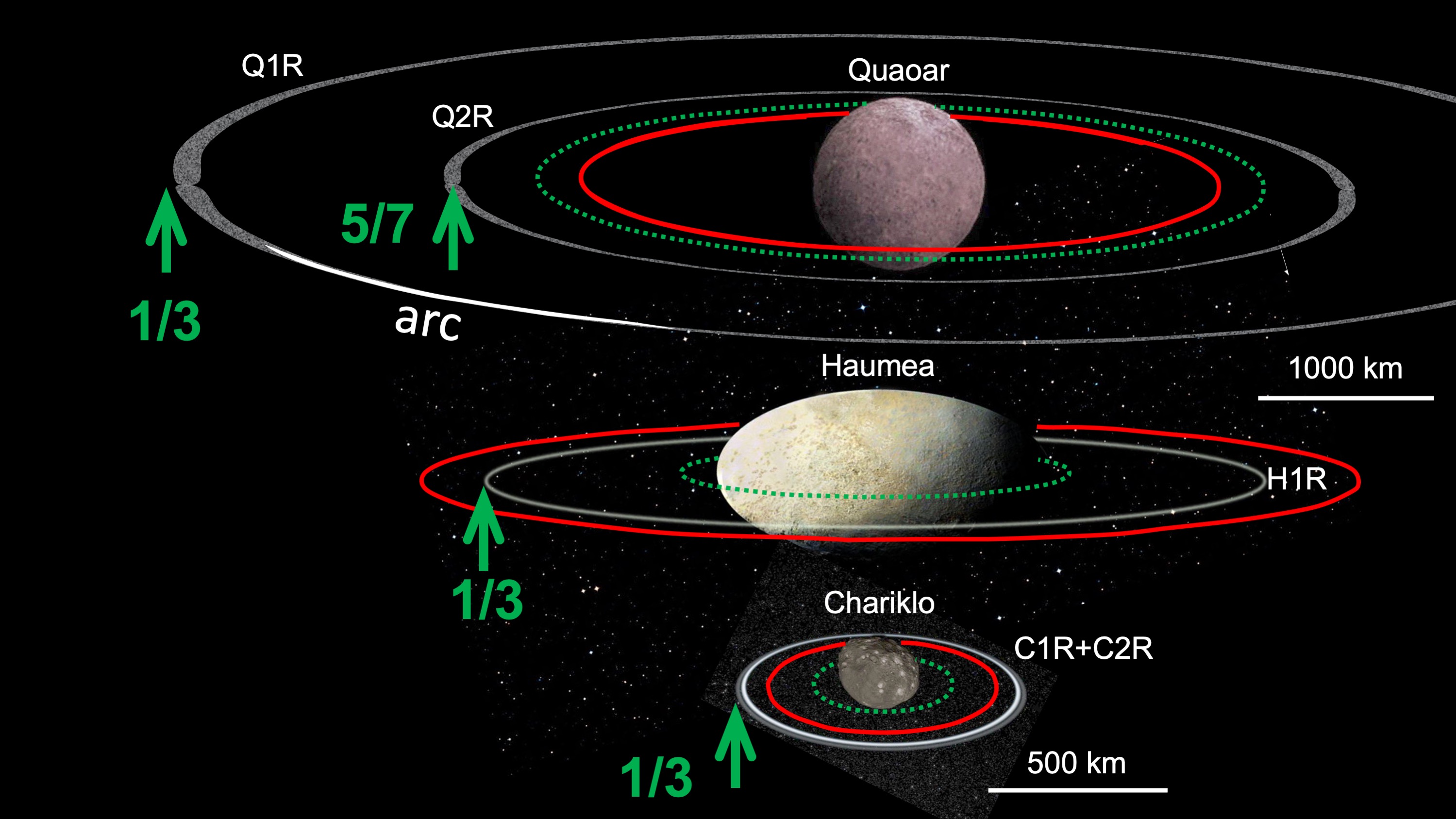}
\caption{%
Artist's views of the rings of Quaoar, Haumea and Chariklo, based on the values in Table~\ref{tab_parameters}.
%
%\bs{%%%
The green dotted lines mark the corotation radii and
%}%%% 
the red solid lines show the Roche limits of each body.
The green arrows indicate the locations of the second-order SORs 1/3 and 5/7.
Quaoar and Haumea are displayed at the same scale, see the 1000-km bar, 
For better viewing, Chariklo's size has been expanded by a factor of two with respect to Quaoar and Haumea, see the 500-km bar.
The widths of the rings have been exaggerated for a better rendering of the figure.
Quaoar's outer ring Q1R is composed of a continuous and tenuous component with typical width of 100~km and normal optical depth of $\tau$=0.5-2\%, and a denser arc of width 5~km and $\tau$=0.25, with an estimated azimuthal extension of 20-70~deg \cite{pere23}. The inner ring Q2R has $\tau$=0.5-1\% \cite{pere23}.
Haumea's ring H1R has a typical width of 70~km and $\tau$=0.1 \cite{orti17}.
Chariklo's inner ring C1R has a typical width of 6~km and average $\tau$=0.4 with almost opaque sharp edges \cite{bera17}. The outer and fainter ring C2R (next to C1R) is less than 1~km in width, with $\tau >$ 0.1 \cite{morg21}.
}%
\label{fig_Chariklo_Haumea_Quaoar}
\end{figure}

Figure~\ref{fig_Chariklo_Haumea_Quaoar} provides a sketch of the rings recently discovered around Chariklo's, Haumea's and Quaoar's rings. Chiron is not included in this picture because the nature of its surrounding material is not yet fully understood. 

Due to their irregular shapes, small bodies have significantly non-axisymmetric gravitational potentials. This creates Spin-Orbit Resonances (SORs), corresponding to commensurabilities between the rotation rate of the body and the mean motion of the ring particles.
From the topology of their phase portraits, it can be shown that only first- and second-order SORs have significant effects on a collisional disk \cite{sica21}. In particular, they cause strong torques that re-arrange the radial distribution of the ring, clearing the synchronous region by pushing material inside and outside this region. This evolution is fast, a few decades for elongated bodies, and a few million years for spherical bodies with topographic features (mountains or craters) of the order of 10~km in height or depth \cite{sica19}
%\bs{%%%
that are actually observed on trans-Neptunian objects \cite{dias17,romm23}.
%}%%%

%%%%%%%%%%%%%%%%%%%%%%%%%%%%%%%%%%%%%%%%%%%%%%%%%%%%%%%%%%%%%%%%%%%%%%%%%%%%%%%%%%%%%%%%%%%%%%%%%
%%%%%%%%%%%%%%%%%%%%%%%%%%%%%%%%%%%%%%%%%%%%%%%%%%%%%%%%%%%%%%%%%%%%%%%%%%%%%%%%%%%%%%%%%%%%%%%%%

Besides this migration effect, resonances can confine ringlets. 
%\bs{%%%
Numerical N-body simulations have shown that small satellites can indeed confine Chariklo's rings through 1st-order mean motion resonances \cite{sick24}. Meanwhile,
%}%%%
all the rings seen so far around small bodies orbit close to the 2nd-order SORs 1/3 and 5/7 with Chariklo, see Fig~\ref{fig_Chariklo_Haumea_Quaoar}. 
This confinement mechanism is backed up by N-body collisional simulations near the 1/3 SOR, where the ring particles complete one revolution when the central body completes three rotations \cite{salo21}.
Thus, the presence of rings near these resonances may be the result of the convergent evolution mentioned in Section~\ref{sec_general}. Even if these rings have different origins, their current locations near the 1/3 (or 5/7) SOR is the result of a common evolutionary path.
% Pas compris - Damya}
% where the orbital period of the particles matches the rotation period of the body. 
The cases of individual objects are now considered in turn.
 
\vspace*{-5pt}
\section{Chariklo and Chiron}
\label{sec_centaurs}

While Chariklo's rings are well defined in terms of orbital elements and optical properties, the situation is more complex for Chiron. Stellar occultations observed between 1995 and 2022 show at some epochs jet-like features around this Centaur object \cite{elli95}, while at other times sharp narrow structures comparable to Chariklo's rings were detected \cite{rupr15,sick20}. 
More recently, these narrow structures were seen as embedded in a diffuse shell \cite{orti23}. The circum-Chiron material could actually be rings in a transient phase of formation, and as such, it may provide valuable information on how rings form around small bodies. 

Compared to other rings around trans-Neptunian objects (Fig.~\ref{fig_Chariklo_Haumea_Quaoar}), Centaurs like Chariklo and Chiron have two distinctive properties: 
(1) they move along unstable orbits over short timescales of millions of years, with close 
encounters\footnote{
%\bs{%%%
In the sense that the objects come to within the planets' Hill spheres.
%}%%%
}
with giant planets \cite{horn04};
(2) Some of the Centaurs have significant cometary activity, and they are massive enough to retain part of the ejected material on bound orbits. These two points are discussed below. 

\subsection{Stability of rings against close encounters}

The stability of rings around Centaurs against
%\bs{%%%
very
%}%%%
close encounters\footnote{
%\bs{%%%
In the sense that the objects come to within a few planetary radii.
%}%%%
}
with giant planets (or other small objects) constrain their ages, and therefore, their origins.
Let us consider a Centaur of mass $M$ with a closest distance approach $\Delta_{\rm CA}$ to a planet of mass $M_{\rm P}$. The ring will be disrupted if its semi-major axis $a$ is larger than the Hill sphere of the Centaur, $R_{\rm H}= \Delta_{\rm CA} (M/3M_{\rm P})^{1/3}$.
This provides the maximum encounter distance 
\begin{equation}
\Delta_{\rm disrupt} \sim a  \left( \frac{3M_{\rm P}}{M} \right)^{1/3}
\label{eq_Delta_disrupt}
\end{equation}
that causes the disruption of the ring.
Considering a giant planet like Uranus ($M_{\rm P}~\sim~10^{26}$~kg) and a Centaur like Chariklo 
($M~\sim~10^{19}$~kg, $a \sim 400$~km), this provides $\Delta_{\rm disrupt}~\sim~130,000$~km, i.e. about five Uranian radii. Similar conclusions are drawn if the planet is Jupiter ($M_{\rm P}~\sim~2~\times~10^{27}$~kg, radius 71,000~km).
Such close encounters have a low probability to occur for Chariklo during its transfer from the trans-Neptunian region to its current orbit, or during subsequent encounters with giant planets \cite{hyod16,arau16,wood17}. However, these encounters may be more frequent for Chiron, implying less stability for its material \cite{arau18}.
%\bs{%%%
We also note that $\Delta_{\rm disrupt}$ is directly proportional to $a$ (Eq.~\ref{eq_Delta_disrupt}), meaning that the larger the ring, the greater the probability that it will be destroyed during an encounter.
%}%%%

%\bs{%%%
Instead of destroying rings, very close encounters might on the contrary be a way to form rings. More precisely, material can be ripped from the object due to the tidal stresses suffered during such encounters, and form later a ring. However,
%}%%%
this requires closest approach distances of the order of 0.4-0.8 planet's Roche limit \cite{hyod16}, which is smaller than $\Delta_{\rm disrupt}$, implying an even lower probability of occurrence.

Besides encounters with giant planets, a Centaur may also encounter other small objects that destabilize its rings. But again such occurrences during the age of the solar system are rare \cite{ikey24}. The same result is obtain by \cite{ikey24} when considering ringed trans-Neptunian objects like Haumea or Quaoar. 

The conclusion here is that encounters of Centaurs with giant planets or small objects are unlikely to trigger the formation of rings or conversely, cause their destruction. In other words, the rings that we see today around Centaurs may have originated a long time ago, when these objects were still moving in the trans-Neptunian region. 

\subsection{Cometary activity of Centaurs}

%\bs{%%%
It is interesting to note that cometary activity is regularly detected around Chiron \cite{meec89,luu90}, while
dense material is detected through occultations in the immediate vicinity of this body, as mentioned earlier. 
In the same vein, the smaller Centaur object Echeclus is also known to have cometary-like episodes. For instance, the March 2006 outburst resulted in an object surrounded by a coma that split from Echeclus \cite{rous08}.
%}%%%
%
The size of this
%\bs{%%%
ejecta
%}%%%
could be as large as about 8 km, corresponding to either a satellite of Echeclus or to a big fragment ejected from its surface. 
%\bs{%%%
In this context, it is worth examining whether or not these materials can constitute a ring-forming pathway.
%}%%%
%
The fate of this material is linked to the escape velocity at the surface of a body,
\begin{equation}
v_{\rm esc}= \sqrt{\frac{8\pi G \rho}{3}}R \sim 
0.75  \left( \frac{\rho}{\rm 
%\red%%%
1000~kg~m^{-3}} \right)^{1/2} \left( \frac{R}{\rm km} \right) 
{\rm ~m~s}^{-1},
\label{equ_escape}
\end{equation}
where $R$ is the radius of the body and $\rho$ is its bulk density.
If $v_{\rm esc}$ is comparable to the velocity $v_{\rm jet}$ of jets at the surface, then the ejected material with not fall back immediately onto the object, nor it will escape to infinity. Instead, it may feed a dusty shell from which a flat disk can emerge at a later stage. Thus, the critical radius that an object must have to keep material in its immediate vicinity without causing its immediate fall on its surface is
\begin{equation}
R_{\sf crit} \sim 
1.3 \left( \frac{v_{\rm jet}}{\rm m~s^{-1}} \right) 
\left( \frac{
%\red%%% 
\rm 1000~kg~m^{-3}}{\rho} \right)^{1/2}
{\rm km}.
\label{equ_crit_radius}
\end{equation}
A typical value of $v_{\rm jet}$ is 100 m~s$^{-1}$ \cite{dels82,teni11}, while densities of Centaurs like Chariklo and Chiron are typically 900-1100 kg~m$^{-3}$ \cite{leiv17,brag23}. This yields $R_{\sf crit} \sim 130$~km, 
%
%\bs{%%%
which corresponds to the sizes of these objects (see \cite{pan16} and Table~\ref{tab_parameters}). In other words, Chiron is large enough to prevent material ejected during outbursts from escaping to infinity, but small enough to prevent this material from immediately falling back onto its surface. This would offer a natural explanation for the presence of rings around Chariklo, if this body had a cometary activity in the 
past\footnote{
%\bs{%%%
Currently, the known active Centaurs have perihelia $q \lesssim 12$~AU \cite{lill24}, while $q=13.1$~AU for Chariklo, so that the body may be a bit too distant to be active now. However, its unstable orbit may have brought it closer to the Sun in the past.
%}%%%
}.
%}%%%
It also explains why Echeclus, with a radius of about 30~km \cite{pere24}, could not retain the ejected material that has been seen in March 2006. 

\vspace*{-5pt}
\section{Haumea}
\label{sec_haumea}

%\bs{%%%
Haumea is a large trans-Neptunian object classified as a dwarf planet. It is a fast rotator with a period of 3.92~h. This causes a very elongated ellipsoidal shape with semi-axes 1161$\times$852$\times$513 km \cite{orti17}. With such a shape, the body is in fact close to the disruption limit under the effects of centrifugal forces. This is illustrated in Fig.~\ref{fig_Chariklo_Haumea_Quaoar} where the corotation radius almost interesects the body.
The apparent orientation and aspect angle of the ring H1R during the discovery occultation in 2017 is consistent with H1R being coplanar with Haumea's main satellite Hi'iaka to within $2 \pm 1.5$ degrees, and also coplanar with Haumea's equator to within $2 \pm 2$ degrees \cite{orti17}.
%}%%%

Currently, Haumea is the only trans-Neptunian object known to be associated with a family \cite{brow07}. The origin of this family is still debated \cite{prou19} and could stem from a
%\bs{%%%
graze-and-merge
%}%%%
collision \cite{lein10} or a binary merging \cite{prou22}. These scenarios would produce a rapidly spinning body that shed material, from which a debris disk and the satellites Namaka and Hi'iaka emerged. The strong SORs induced by the large elongation of Haumea then make any orbit inside a radius of about 1900~km highly unstable \cite{ribe23}. The 1/3 SOR would in fact be the first resonance that is at the same time weak enough not to disrupt a ring, but also not too far away from the dwarf planet to prevent accretion into satellites.

%\bs{%%%
Thus, contrarily to Chariklo's rings that may be produced by the lofting of dust particles ejected from its surface, the origin of Haumea's ring is consistently with a catastrophic impact scenario.
%}%%%

\vspace*{-5pt}
\section{Quaoar}
\label{sec_quaoar}

Quaoar is the only body of the solar system known so far to host dense collisional rings well outside its Roche limit (excluding very tenuous dusty rings for which this limit is irrelevant).
The radii of Q1R and Q2R correspond respectively to about 7.4 and 4.6 Quaoar's radius (Table~\ref{tab_parameters}). Assuming again icy ring particles with density 400~kg~m$^{-3}$, the Roche limit lies at roughly 3.2 Quaoar's radius. In these conditions, and unless the ring material 
%\bs{%%%
has an unrealistically low density of 30~kg~m$^{-3}$ \cite{morg23},
%}%%% 
the rings should accrete into satellites and thus disappear on time scales as short as a few months. 
The question therefore arises as to whether the rings derive from a completely different pathway compared to the other ring systems, or whether they originate from the same basic processes, the only difference being that they can survive at unusually large distances from the body.

This paradox can be resolved by realizing that collisions between ring particles may be more elastic than previously estimated. Laboratory experiments with icy particles at $T=210$~K show that particles at the distance of Q1R should collide with low rebound coefficients of less than about 0.4 \cite{brid84}. Conversely, measurements made in a colder environment (T=125~K) result in more elastic collisions with rebound coefficients as high as 0.7 \cite{hatz88}. Quaoar's rings have an even lower temperature (about 30~K), possibly resulting in even more elastic collisions.    

Local N-body simulations
%\bs{%%%
\cite{morg23}
%}%%% 
then show that in the less elastic model at 210~K, Q1R very rapidly accretes into small moons, while in the colder case at 125~K, the post-impact relative velocity of the particle maintains itself above their mutual escape velocity, preventing accretion even for particles as dense as 5000~kg~m$^{-3}$.
%\bs{%%%
Thus, while porous particles with densities as low as 400~kg~m$^{-3}$ (as in Saturn's rings) may result in rather inelastic collisions, denser and harder icy particles with more elastic collisions may escape accretion at the level of the Q1R and Q2R rings.
%}%%%

Another point to consider is that both Q1R and Q2R are close to SORs (see Section~\ref{sec_resonances} and Fig.~\ref{fig_Chariklo_Haumea_Quaoar}), and for Q1R, to the
%\bs{%%%
concomitant
%}%%%
effect of the 6/1 mean motion resonance with the satellite Weywot. These resonances might help to maintain the relative velocities between ring particles high enough to prevent accretion. 
In this context, and whatever its origin is, an initial broadly distributed collisional disk around Quaoar may rapidly accrete into small moons, except near resonances that act both as a confining mechanism and as the source of velocity dispersion that prevents accretion. A possible test of this model is to obtain high resolution images of Quaoar's environment
%\bs{%%%
to reveal small (say 10 km) satellites interspersed with the rings. However, detecting such small objects remains challenging. At the geocentric Quaoar distance (about 42 AU), they are angularly separated from the primary by less than $\sim$0.13~arcsec, while being up to 10~magnitudes fainter than Quaoar. This is a task that can be better tackled using future 30m-class telescopes. Another approach is to detect these satellites through stellar occultations. However, this can be achieved only on serendipitous basis with a low chance of success.
%}%%%

The improvement of Weywot's orbital elements will also help understanding the strongly inhomogeneous azimuthal structure of Q1R, reminiscent of Neptune's ring arcs. Such structures can be maintained through corotation resonances that appear if the orbit of the 
%\bs{%%%
perturber
%}%%% 
(here Weywot) is eccentric or if co-orbital satellites create Lagrange-type corotation sites along Q1R.

An interesting question related to the origin of Quaoar's rings is whether they may have formed when Weywot was much closer to the primary. In this scenario, the rings would have first formed inside Quaoar's Roche limit, then they were confined at the 5/7 and 1/3 SORs, and finally they migrated outwards as Weywot receded from Quaoar due to tidal effects. 

\vspace*{-5pt}
\section{Conclusions}

The rings discovered so far in the solar system come in very different sizes and they surround very different bodies. It is, thus, difficult to imagine a unique, generic framework to explain their formations.
As pointed out in Section~\ref{sec_general}, another difficulty arises as the ring features that we observe now may be convergent, so that common short-lived features may mask different origins.
 
Here, we have presented several scenarios for each of the aforementioned bodies. 
For instance, the tenuous and dusty rings of Jupiter can be explained by the meteoroid bombardment of small inner Jovian satellites that produces the observed dust. Even though these rings reached a stationary regime over the long term, a given dust particle has a much shorter lifetime.

%\bs{%%%
Uranus' and Neptune's rings are best explained by the continuous formation and fragmentation of small parent satellites during cycles of destructive collisions and re-accretion processes.
%}%%% 
% that move on unstable orbits. 
These satellites may be the remnants of a primordial disk that surrounded the planets at the very beginning of their formation.
%\bs{%%%
We note in passing that Triton's orbital decay due to tidal effects may end up with the destruction of the satellite in some 1.4 Gyr \cite{chyb89}. If this is the case, Neptune might host in the future the most spectacular ring of the solar system. As it will be retrograde, future astronomers will easily recognize its exogenous origin.
%}%%%

In spite of the large amount of data returned by the Voyager and Cassini spacecraft, the origin of Saturn's rings remains enigmatic. Their formation and age continue to spark scientific inquiry and debate. They could be primordial, having slowly spread and formed small satellites near their outer edge. Conversely, they may be the recent result of the breakup of a satellite that spiraled inwards and got disrupted when entering Saturn's Roche zone. Each proposed model offers insights into different aspects of their formation, challenging our understanding of planetary ring systems and their dynamic evolution over time. As of now, determining the age of Saturn's rings seems impossible.
%However, a future mission dedicated to studying the rings and their interaction with the gas giant over an extended period could provide the necessary data to achieve this goal.
%\pd{Bruno to Maryame? All?: I don't know how to respond to Reviewer \#2's comment: ``Which measurements from a future mission could determine the age of Saturn's rings?"}
%\bs{%%%
In particular, it is still difficult to determine which critical observable can discriminate once and for all the “old/young” dilemma concerning Saturn's rings \cite{char18}. Of course, in situ (or sample-return) analysis of the surface and interior of the ring particles would be of great help, but it will be several decades before this is possible.
%}%%%

Concerning rings around small objects, it seems that some generic processes independent of their origins are at work. In particular, the SORs associated with the non-axisymmetric potentials of the bodies can rapidly clear the synchronous orbit, while some specific SORs (in particular the 1/3 resonance) confine the disk material into narrow ringlets.

Turning to Centaurs, cometary activity could eject material from their surfaces and feed dense shells that evolve to form a collisional disk. 
As for the dwarf planet Haumea, the ring may be the result of material shed from the rapidly rotating body after a grazing impact or a binary merging. Due to the highly elongated shape of the body, very strong resonances cause the rapid clearing of the immediate vicinity of Haumea due to strongly perturbed orbits inside a radius of $\sim$1900~km. 
%{\color{red} DS- the transition here is not natural, I will think of something but f someone has an idea, please go ahead. - Then,} 
% When moving outwards, the initial disk surrounding Haumea would accrete in satellites, explaining for instance the presence of the satellites Namaka and Hi'iaka.
Then, only the outermost 2nd-order resonance 1/3 remains a viable scenario to confine a ring. 

Quaoar's rings are more puzzling. They are narrow and dense, as such they remind
%\bs{%%%
us
%}%%%
the rings of Uranus and Neptune. However, they are well beyond Quaoar's Roche limit. This challenges the hitherto accepted idea that dense rings should not exist outside this limit. The answer to this paradox could be that the low temperatures prevailing at Quaoar's heliocentric distance make collisions more elastic than was previously estimated, thus preventing accretion. However, this also requires
%\bs{%%%
maintaining
%}%%%
a high enough velocity dispersion between the particles in the rings through the action of resonances.

%\bs{%%%
Progress in the knowledge of these rings could be made by detecting optically thin rings. This is a challenging task to perform from Earth, as this faint material remains angularly very close to the object, typically less than 0.2~arcsec. It is then overwhelmed by the point spread function of the much brighter central body, even using large ground-based instruments with adaptive optics systems or space telescopes \cite{sica15}. The NASA New Horizons LORRI camera could search faint material around Quaoar or Haumea in the forward scattering geometry, where dusty material is better seen. However, it remains to be seen if the sensitivity and resolution of LORRI would permit such detections.
%}%%%

A point not addressed here is the fact that rings are seen only in the outer solar system. Hundreds of asteroids were scrutinized using stellar occultations, and high resolution images taken from Earth or from spacecraft have not revealed the existence of rings.
%
%\bs{%%%
This situation may change on a relatively short timescale. The martian satellite Phobos is expected spiral and enter Mars' Roche limit within a few tens of million years \cite{efro07}. So, it may be that the inner solar system hosts at that time a new ring system.
%}%%%
%
%\bs{%%%
Returning to the initial question,
%}%%%
a possible explanation is that only rings composed of water ice can exist \cite{hedm15}. This is in line with the predominant presence of water ice found in Saturn's \cite{cuzz18} and Chariklo's \cite{duff14} rings. In other words, water ice would have special properties that allow the formation of countless meter-sized ring particles, something that rocky material cannot achieve. 
However, for the time being, the specific properties of water ice that allow such behavior have not been explained. The answer to this question could represent a major step forward for understanding the origin of rings.

\ack{The authors thank Luke Dones and an anonymous referee for comments and corrections that have improved this paper.
R.S. acknowledges support by the DFG German Research Foundation (project 446102036).}

\bibliographystyle{RS}		% Native  bibliographystyle
\bibliography{biblio}

%%%%%%%%%% Insert bibliography here %%%%%%%%%%%%%%

% \begin{thebibliography}{9}

% \bibitem{1} Allwood JM, Cullen JM. 2011 \textit{Sustainable materials:  with both eyes open}.
% Cambridge, UK: UIT Cambridge. See \href{http://www.withbotheyesopen.com}{http://www.withbotheyesopen.com}.

% \bibitem{2}  MacKay DJC. 2008  \textit{Sustainable energy:  without the hot air}.
% Cambridge, UK: UIT Cambridge. See \href{http://www.withouthotair.com}{http://www.withouthotair.com}.

% \bibitem{3} Gallman PG. 2011  \textit{Green alternatives and national energy strategy: the facts
 % behind the headlines}.  Baltimore,\ MD: Johns Hopkins University Press.

% \bibitem{4} MacKay DJC. 2013.  Solar energy in the context of energy use, energy transportation, and
% energy storage. \textit{Proc. R. Soc. A} \textbf{371}.

% \end{thebibliography}

\end{document}